# Orbital wave in the Raman scattering cross section of LaMnO$_3$

Purevdorj Munkhbaatar[1,2] and Kim Myung-Whun[1,3,*]

[1]*Institute of Photonics and Information Technology, Chonbuk National University, Jeonju 54896, Republic of Korea*

[2]*Department of Physics, National University of Mongolia, Ulaanbaatar 210646, Mongolia*

[3]*Department of Physics, Chonbuk National University, Jeonju 54896, Republic of Korea*

[*]*corresponding author: mwkim@chonbuk.ac.kr*

We calculated the polarization-dependent Raman scattering cross section spectra of LaMnO$_3$ below the A-type magnetic ordering temperature. Two strong peaks appear around the MnO$_6$ octahedra stretching phonon frequency. One mode shows A$_g$ symmetry, while the other mode shows B$_g$ symmetry. We found that the A$_g$ symmetry peak is a Jahn–Teller phonon coupled to the orbital wave and the B$_g$ symmetry peak is an orbital wave coupled to a Q$_2$ phonon mode via the Jahn–Teller electron-phonon coupling.

When light causes photo-excitation in a solid material, part of the incident photon energy is absorbed with the generation of Raman mode vibrations in the material. From a classical point of view, Raman mode vibration occurs because the polarizability of the charges surrounding an ion acts as a spring constant [1]. This can be a sufficient explanation for many materials, but not for orbitally degenerate Mott insulators such as $LaMnO_3$.

Imagine the Raman process occurring on the microscopic scale in $LaMnO_3$. A Mn–3d electron absorbs a photon and jumps to a neighboring Mn-site. A hole-doublon pair is created by the inter-site transition. The process of creation and destruction of the hole-doublon pair can be explained approximately by the dynamic polarizability change because the pair causes the spatial variation of effective charge density; however, there is an important part missing from this explanation. The outermost electron orbitals of $LaMnO_3$ are doubly degenerate, and one of these degenerate orbitals is occupied by one Mn–3d electron. $LaMnO_3$ is a Mott insulator, and the interaction of the outermost electrons between the lattice sites must occur between electrons that occupy different orbitals. This will minimize Hund energy and Coulomb potential energy, according to the Pauli exclusion principle [2,3]. The Raman mode occurring in this process is a quantum mechanical phenomenon that cannot be explained by a simple polarizability change.

Saitoh *et al.*, for the first time, paid attention to the abnormal phenomenon exhibiting the quantum mechanical effect in the Raman spectra. Saitoh *et al.*, emphasized the weak signals around 150 meV in the Raman spectrum of $LaMnO_3$ [4]. They claimed that a weak signal near 150 meV is a new elementary excitation that differs from the simple phonon. However, other researchers suspected that the weak signal is due to the first overtones of fundamental phonon modes [5]. Until recently, studies on the Raman scattering signal of $LaMnO_3$ remained controversial despite the accumulation of precise spectroscopic studies for high-quality crystals of $LaMnO_3$ and its relatives [6–8]. The origin of the high frequency Raman mode that Saitoh *et al.*, highlighted is an important question, but a more important question would be how we understand the fundamental modes of a Raman scattering spectrum in orbitally degenerate Mott insulators. Thus far, this question has been rarely considered.

In this Letter, we present the Raman scattering cross section spectra for an ideal LaMnO$_3$ with one electron filled in the doubly degenerate Mn 3d e$_g$ orbitals at low temperature. We investigate how the scattering cross section varies depending on the light polarizations and compare the calculation results with previous experiments. In particular, we focus on the discussion of the origin of a strong peak at 611 cm$^{-1}$.

We adopted Brink's Hamiltonian to calculate the energy dispersion curve and the Raman scattering spectra [9];

$$H = H_{orb} + H_{e-ph} + H_{ph}. \qquad (1)$$

The spin degree of freedom is frozen. The first term describes the super-exchange interaction between the orbitals;

$$H_{orb} = J \sum_{\langle ij \rangle_\Gamma} T_i^\Gamma T_j^\Gamma, \qquad (2)$$

where the sum is over neighboring sites $\langle ij \rangle$ along the $\Gamma = a, b$ crystal axis and $J$ is exchange coupling. The orbital operators, $T_i^\Gamma$, can be expressed in terms of the pseudo-spin operator ($\tau$): $T_i^{a/b} = (\tau_i^z \pm \sqrt{3}\tau_i^x)/2$, and the plus (minus) sign corresponds to $a$ ($b$) axis. The second term of Eq. (1) represents the Jahn–Teller electron-phonon coupling;

$$H_{e-ph} = g \sum_i (\tau_i^z Q_{3i} + \tau_i^x Q_{2i}), \qquad (3)$$

where $g$ is the electron-phonon coupling constant and $Q_{2/3,i} (= a_{2/3,i}^\dagger + a_{2/3,i})$ are phonon operators of Jahn–Teller modes with e$_g$ symmetry. The third term of Eq. (1) is the phonon contribution;

$$H_{ph} = \omega_0 \sum_i (a_{2i}^\dagger a_{2i} + a_{3i}^\dagger a_{3i}) + \omega_1 \sum_{\langle ij \rangle_\Gamma} Q_i^\Gamma Q_j^\Gamma, \qquad (4)$$

where $\omega_0$ is the local Jahn–Teller phonon frequency for the $Q_2$ and $Q_3$ modes. Here, $\omega_1$ is the nearest-neighbor coupling between the phonons. The $Q_i^{a/b}(= (Q_{3i} \pm \sqrt{3}Q_{2i})/2)$ represents the coupled Jahn–Teller mode along a crystal axis, $a$ or $b$.

We assumed antiferromagnetic-ordered four Mn-ion $e_g$ orbitals forming two sublattices (A and B) as the unit cell shown in the inset of Fig. 1(a). To describe the orbital part, we introduce an operator, $\tilde{\tau}_{i\lambda} = (\tilde{\tau}_{i\lambda}^z, \tilde{\tau}_{i\lambda}^x) = (\tau_i^z \cos\theta_\lambda - \tau_i^x \sin\theta_\lambda, \tau_i^x \cos\theta_\lambda + \tau_i^z \sin\theta_\lambda)$, for the $\lambda$ sublattice. We rewrite the localized pseudospin operators in terms of Holstein Boson operators ($a$);

$$\tilde{\tau}_i^x = (a_{1i} + a_{1i}^\dagger)/2, \quad \tilde{\tau}_i^z = (1/2 - a_{1i}^\dagger a_{1i}). \qquad (5)$$

We assumed that the angle on each sublattice is $\theta_{A/B} = \pm\theta$. After inserting the expansion into the full Hamiltonian, Eq. (1) becomes,

$$H = \omega \sum_i a_{1i}^\dagger a_{1i} + J/4 \sum_{\langle ij \rangle} \tilde{\tau}_i^x \tilde{\tau}_j^x + g \sum_i \tilde{\tau}_i^x (\cos(\theta)Q_{2i} \pm \sin(\theta)Q_{3i}) + \omega_0 \sum_i (a_{2i}^\dagger a_{2i} + a_{3i}^\dagger a_{3i}) + \omega_1 \sum_{\langle ij \rangle_\Gamma} Q_i^\Gamma Q_j^\Gamma. \qquad (6)$$

The first term in Eq. (6) is the phenomenological term representing the localized orbiton. Here, $\omega$ includes the superexchange interaction and the static Jahn–Teller deformation energy. We assumed that $\omega$ is equal to $\omega_0$ (the bare phonon frequency of Jahn–Teller phonon $Q_2$ mode) under the influence of the strong Jahn–Teller type electron-phonon coupling. Therefore, the local orbital motion should be locked to the local lattice motion. The plus (minus) sign is for A (B) sublattice. In Eq. (6) we collected only the quadratic terms of Boson operators for simplicity [10]. The quadratic Hamiltonian is diagonalized by the Bogoliubov transformation. The transformed Hamiltonian is $\tilde{H} = \sum_{\mu\,k} E_\mu(\mathbf{k})\, \alpha_\mu^\dagger(\mathbf{k})\alpha_\mu(\mathbf{k})$. The energy eigenvalues of the transformed Hamiltonian are shown in the energy dispersion curves.

Next, we determined the Raman scattering cross section based on the calculation of the superexchange interaction between neighboring orbital operators by using the Shastry–Shraiman method [11];

$$R(\Omega) = \sum_f \delta(\Omega - (E_f - E_g)) \left| \sum_{\langle ij \rangle_\Gamma} e_{f\Gamma} e_{i\Gamma} \langle f | (1/2 - T_i^\Gamma)(1/2 - T_j^\Gamma) | g \rangle \right|^2. \qquad (7)$$

Here, $\Omega$ $(= \omega_i - \omega_f)$ is the Raman scattering energy shift. $E_{g(f)}$ represents the energy of the ground (final) state and $e_{f(i)\Gamma}$ is the scattering (incident) photon polarization vector along $\Gamma$ axis.

We obtained the fundamental mode scattering cross section by collecting the first-order terms of Holstein bosons:

$$R_1(\Omega) = \sum_f \delta\left(\Omega - (E_f - E_g)\right) \left| \sum_\Gamma e_{f\Gamma} e_{i\Gamma} \langle f | \sum_i \rho_\Gamma{}^i (a_{1i} + a_{1i}^\dagger) | g \rangle \right|^2, \qquad (8)$$

with $\rho_{a/b}{}^i = \mp\cos[\pi/6 \mp \theta_\lambda](1 - \sin[\pi/6 \pm \theta_\lambda])$ if $i \in \lambda$ ($\lambda$ = A or B) sublattice. After the Bogoliubov transformation, Eq. (8) becomes

$$\sum_\mu \delta(\Omega - E_\mu(\mathbf{0})) \left| \sum_\nu e_{f\Gamma} e_{i\Gamma} \rho_\Gamma{}^\nu (V_{\nu\mu}(0) + W_{\nu\mu}^*(0)) \right|^2. \qquad (9)$$

Here, $V_{\nu\mu}(\mathbf{k})$ and $W_{\nu\mu}(\mathbf{k})$ are Bogoliubov transformation coefficients connecting the boson operator for the $\nu$th orbiton in a unit cell to that of the $\mu$th eigenmode with the energy, $E_\mu(\mathbf{k})$, as

$$a_{1\nu}^\dagger(k) = V_{\nu\mu}(k) \alpha_\mu^\dagger(k) + W_{\nu\mu}(k) \alpha_\mu(-k). \qquad (10)$$

We obtain the scattering cross section of the overtone mode by collecting the second-order terms of Holstein bosons:

$R_2(\Omega) =$

$\sum_f \delta\left(\Omega - (E_f - E_g)\right) \left| \sum_\Gamma e_{f\Gamma} e_{i\Gamma} \langle f | 4 \sum_i \sigma_\Gamma{}^i a_{1i}^\dagger a_{1i} + \rho \sum_{\langle ij \rangle_\Gamma} (a_{1i} + a_{1i}^\dagger)(a_{1j} + a_{1j}^\dagger) | g \rangle \right|^2,$

(11)

with $\rho = \cos[\frac{\pi}{6} - \theta]\cos[\frac{\pi}{6} + \theta]$, and $\sigma_{a/b}{}^i = \sin[\pi/6 \mp \theta_\lambda](1 - \sin[\pi/6 \pm \theta_\lambda])$ if $i \in \lambda$ ($\lambda$ = A or B) sublattice. By the Bogoliubov transformation, Eq. (11) becomes;

$$\sim \sum_{k\mu\mu'} \delta(\Omega - E_\mu(k) - E_{\mu'}(-k)) \left| \sum_{\langle\nu\nu'\rangle_\Gamma} e_{f\Gamma} e_{i\Gamma} \rho \left(V_{\nu\mu}(k) + W_{\nu\mu}^*(k)\right)\left(V_{\nu'\mu'}(-k) + W_{\nu'\mu'}^*(-k)\right)(1 + \cos(k_\Gamma)) + 4 \sum_\nu e_{f\Gamma} e_{i\Gamma} \sigma_\Gamma^\nu V_{\nu\mu}(k) W_{\nu\mu'}^*(-k) \right|^2. \tag{12}$$

We chose 20×20 lattice cells for the numerical calculations. Figure 1 shows the calculation results for Raman scattering cross sections at the $\Gamma$–point for the fundamental mode. Figure 1 (a–c) shows the cases where the polarization of incident light and scattered light is ($x'x'$), ($x'y'$), and ($xx$), respectively. As shown in the inset in Fig. 1(a), $x$ and $y$ are the two orthogonal bonding directions of Mn–O–Mn, and $x'$ and $y'$ are diagonal directions, i.e., $x+y=x'$ ($x-y=y'$). In the ($xy$) polarization, no modes are active.

Before looking at the calculation results, we reviewed some of the points observed in previous experiments: (1) when the polarization of the incident light is the same as the polarization of the scattered light and the direction of the electric field is parallel to the crystal axis [for example, ($xx$) polarization], Raman experiments using incident light with a wavelength of 632 nm show two strong peaks at 496 cm$^{-1}$ and 611 cm$^{-1}$. The peak at 611 cm$^{-1}$ is more intense than the peak at 496 cm$^{-1}$ [6]. (2) The peak at 496 cm$^{-1}$ has $A_g$ polarization symmetry and the peak at 611 cm$^{-1}$ has $B_g$ polarization symmetry [6]. (3) The peak center frequency is slightly different (~ 5 cm$^{-1}$) depending on the polarizations. For example, at 5 K, the peak center frequency near 496 cm$^{-1}$ is 500 cm$^{-1}$ for the ($x'x'$) polarization and 495 cm$^{-1}$ for ($x'y'$) polarization [12]. (4) There are three peaks near 1200 cm$^{-1}$, whose frequencies are approximately twice as large as 496 cm$^{-1}$ and 611 cm$^{-1}$ [4]. (5) The intensity of the three peaks near 1200 cm$^{-1}$ is approximately 20% of the intensity of the two peaks near 550 cm$^{-1}$ [6]. Our calculation results, which can reproduce all of these experimental results, have not yet been reported.

The energy parameters used to obtain Fig. 1 are $\omega = \omega_0$, $\omega_1 = -0.1\omega_0$, $J = 0.5\omega_0$, $g = 0.2\omega_0$, and $\theta = \pi/2$. Assuming that $\omega_0$ is 557 cm$^{-1}$, the calculated spectra are consistent with all the five experimental observations listed in the previous paragraph. In Fig. 1(a), there are two strong peaks in the vicinity of $\Omega/\omega_0 = 1$ for $(xx)$ polarization; one at $0.9\omega_0$ and the other at $1.1\omega_0$ ($\Omega$ is the measurement frequency). The peak intensity at $1.1\omega_0$ is more intense than that at $0.9\omega_0$. Figures 1(b) and 1(c) show the $(x'x')$ polarization and $(x'y')$ polarization, respectively. The $(x'x')$ polarization spectrum shows a $0.9\omega_0$ peak while the $(x'y')$ polarization spectrum shows a $1.1\omega_0$ peak. The $0.9\omega_0$ peak has A$_g$ symmetry and the $1.1\omega_0$ peak has B$_g$ symmetry, which agrees well with the experiment. However, as can be seen from the experimental results, the symmetry of the two peaks is not perfect. The $(x'x')$ polarization shows a weak peak at $1.1\omega_0$ and the $(x'y')$ polarization shows a weak peak at $0.9\omega_0$ [12]. The cause of the incomplete polarization dependence of the experiment could be an incomplete polarizer or an imperfection of the sample such as crystal twinning. However, our results suggest that even in the ideal experimental situation, the $1.1\omega_0$ peak appears in A$_g$ symmetry and the $0.9\omega_0$ peak appears in B$_g$ symmetry if the Jahn–Teller electron-phonon coupling value is larger than 0. The larger the $g$ value, the greater the intensity of the forbidden symmetry peak. The left inset of Fig. 1(b) and the left inset of Fig. 1(c) show the polarization dependence of the $1.1\omega_0$ peak and $0.9\omega_0$ peak, respectively. It can be seen that the center frequency of the peak is slightly different (~ $0.01\omega_0$ corresponding to 5.5 cm$^{-1}$) for each polarized light, and it is consistent with the experimental results [12].

The calculation also shows three relatively weak peaks at $\Omega/\omega_0 = 1.8$, 2, and 2.2. These are first-overtone modes. The scattering intensities of the overtone peaks are almost 20% of the fundamental modes. The polarization dependence of the overtone modes is different from that of the fundamental modes. The intensity of $(x'x')$ polarization is slightly weaker than that of $(xx)$. These results are consistent with the experiments conducted with a 632 nm laser [6,7,12]. This is because our calculation assumes the Mott–Hubbard transition at 2 eV as the intermediate state excitation. As shown in the right inset of Fig. 1(c), the calculation shows that the frequencies of the high-energy Raman band cannot be reproduced by simple addition or multiplication of the fundamental mode frequencies. This is also consistent with the experiment [4].

We investigated the dispersion of energy eigenmodes, $E_\mu(k)$, for several different energy parameters to understand the origin of the Raman mode peaks. The important parameters are $J$, $g$, and $\omega_1$. Figure 2 shows how these three parameters play a role in the scattering cross section. Figure 2 (a–c) shows the results when $J = 0.5\omega_0$, $g = 0$, and $\omega_1 = 0$. The straight line at $E_\mu(k)/\omega_0 = 1$ represents the dispersion of the Jahn–Teller phonon. The curved lines correspond to the orbital wave modes. $J$ is responsible for the energy bandwidth of the orbiton dispersion. We show the fundamental mode scattering (near $\Omega/\omega_0 = 1.0$) and first overtone scattering (near $\Omega/\omega_0 = 2.0$) signals in Fig. 2(b) and 2(c), respectively. Dispersion curve analysis shows that the peak at $\Omega/\omega_0 = 0.9$ in Fig. 2(b) is a pure orbiton with $A_g$ symmetry, the peak at $\Omega/\omega_0 = 1.1$ is a pure orbiton with $B_g$ symmetry. Both peaks are generated by the super-exchange type interaction between neighboring Mn $e_g$ orbitals. These peaks correspond to the orbital wave or orbiton proposed by Saitoh et al [4]. There is a broad single peak around $\Omega/\omega_0 = 2.0$ and the intensities of those peaks are slightly different for different polarizations. Figure 2(d–f) shows the results for $\omega = \omega_0$, $J = 0.5\omega_0$, $g = 0.2\omega_0$, and $\omega_1 = 0$. Here, $g$ indicates the phonon-orbital coupling by the Jahn–Teller mechanism. The dispersion curves shown in Fig. 2(a) are split and the energy difference between the main curves corresponds to $g$. The low energy $[E_\mu(\mathbf{k})/\omega_0 < 1]$ curves in the dispersion, correspond primarily to the $Q_2$ mode Jahn–Teller phonon while the straight line at $E_\mu(\mathbf{k})/\omega_0 = 1$ represents the dispersion of the $Q_3$ mode phonon. The curve at high energy $[E_\mu(\mathbf{k})/\omega_0 > 1]$ corresponds to the orbital wave mode. In Fig. 2(e), four scattering peaks appear near $\Omega/\omega_0 = 1.0$. Two more intense peaks are main peaks and the weaker peaks are the satellites. Phonon and orbiton characteristics are mixed in the peaks and the intensities are similar at different polarizations. In Fig. 2(f), three peaks appear near $\Omega/\omega_0 = 2.0$ and the intensities at different polarizations are similar. Any change in $g$ affects the energy difference of the two peaks and the intensity ratio between fundamental and overtone peaks. We chose the $g$ value to meet the value of 20%—the experimentally observed ratio between the fundamental modes and the first overtone modes [6].

Figure 2(g–i) shows the results for $\omega = \omega_0$, $J = 0.5\omega_0$, $g = 0.2\omega_0$, and $\omega_1 = -0.1\omega_0$, as the same as those of Fig. 1. Here, $\omega_1$ represents the phonon-phonon interaction between the local phonons. The shape of the curve is very complicated due to the non-zero value of $\omega_1$. However, the scattering peaks are not much different from those of Fig. 2(b), which means that the $A_g$ symmetry peak is from a phonon and the $B_g$ symmetry peak is from an orbiton. However, because $g$ is finite, the orbiton character is mixed in the $A_g$ symmetry peak and that of the phonon is mixed in the $B_g$ symmetry peak. The peak observed at 611 cm$^{-1}$ in the experiment is therefore, the orbiton coupled to the phonon through the Jahn–Teller mechanism. Among the three peaks near $\Omega/\omega_0 = 2.0$, the highest energy peaks are due to a two-orbiton excitation. Middle energy peaks are due to a $Q_3$ phonon + orbiton excitation. The lowest energy peak is due to a two-$Q_2$-phonon excitation.

The last puzzle is the intensity change of each peak depending on the wavelength of incident light. According to the experiment of Krüger et al., when (*xx*) polarized light with a wavelength of 632 nm is incident, peaks at 496 cm$^{-1}$ and 611 cm$^{-1}$ are clearly observed. However, the incident light with a wavelength of about 276 nm still results in the sharp peak at 496 cm$^{-1}$, but the incident light seems not to excite the peak at 611 cm$^{-1}$ [6]. If we assign the peak at 611 cm$^{-1}$ as a simple Mn–O bond stretching phonon mode, we cannot explain the light wavelength dependence of the peak intensity.

The absorption spectrum of LaMnO$_3$ shows a weak and a broad peak near 2 eV (about 620 nm) and a much stronger absorption above approximately 3 eV (approximately 410 nm) [2]. Previous studies have shown that the absorption of 2 eV is mainly due to the Mn 3d → Mn 3d inter-atomic transition, whereas the absorption above 3 eV is mainly due to the O 2p → Mn 3d transition [2,3]. When the incident photon energy is larger than 3 eV, the O 2p → Mn 3d charge transfer transition is dominant, and the Raman spectra is dominated by the Franck–Condon process [13]. An incident photon (> 3 eV) excites an electron in the initial state of O 2p orbital [Fig. 3(a)] and the electron transits to the neighboring Mn 3d orbital leaving a hole in the O 2p orbital. A hole-doublon state is formed [Fig. 3(b)] and the hole-doublon becomes localized due to the neighboring Mn $e_g$ orbital configurations (Mn–Mn inter-site hopping is suppressed). Because of the orbital configuration, the most probable path for the excited electron is the return path to the initial O 2p orbital and finally a Mn–O phonon is generated [Fig. 3(c)]. Even after the transition process is complete, the long-range collective orbital configuration cannot be changed very much. As a result, the orbiton cannot be excited, while the Mn–O phonon can be excited. This model explains why the 611 cm$^{-1}$ peak is excited by 632 nm light but not by 276 nm light.

In conclusion, our calculations show that the peak at 611 cm$^{-1}$ is an orbiton mode coupled with a phonon. We found that the three peaks near 1200 cm$^{-1}$ (approximately 150 meV), which have been assigned as orbitons in previous studies, do not arise from single orbitons, but are a two-phonon peak, a phonon-orbiton synthesized peak, and a two-orbiton peak.

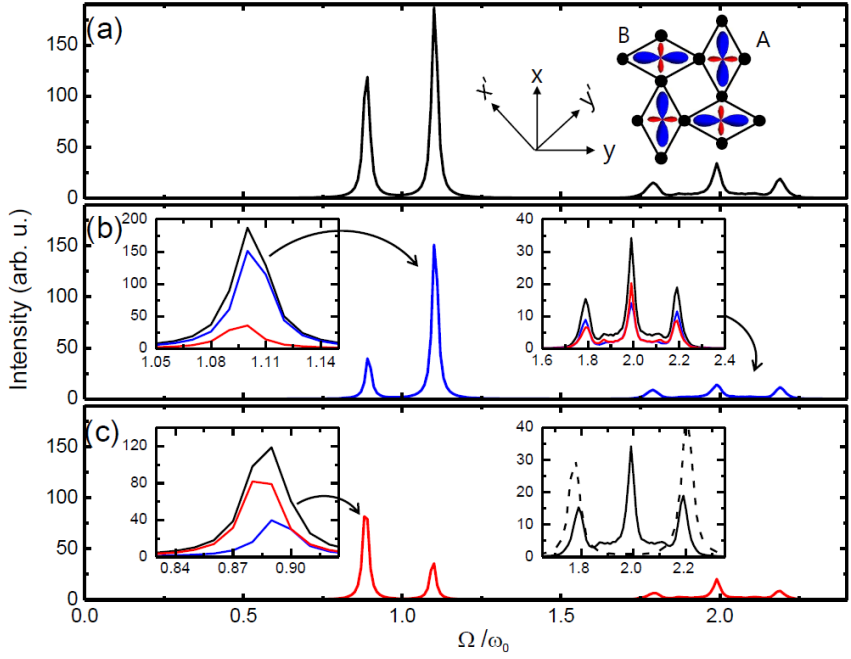

**FIG. 1** (color online) Raman scattering cross section spectra for different light polarizations: (a) The spectrum for the (*xx*)-polarization. Inset shows the unit cell lattice, orbital configurations, and the polarization direction with respect to the lattice. (b) The spectrum for the (*x'x'*)-polarization. Left inset shows the polarization dependence of the peak at $1.1\,\omega_0$ (solid line: (*xx*), dashed line (*x'x'*), and dotted line (*x'y'*)). Right inset shows the polarization dependence of the three peaks near $\Omega/\omega_0 = 2.0$. (c) The spectrum for the (*x'y'*)-polarization. Left inset shows the polarization dependence of the peak at $0.9\omega_0$ (solid line: (*xx*), dashed line (*x'x'*), and dotted line (*x'y'*)). Right inset compares the energy of the peaks near $\Omega/\omega_0 = 1.0$ with that of the three peaks near $\Omega/\omega_0 = 2.0$.

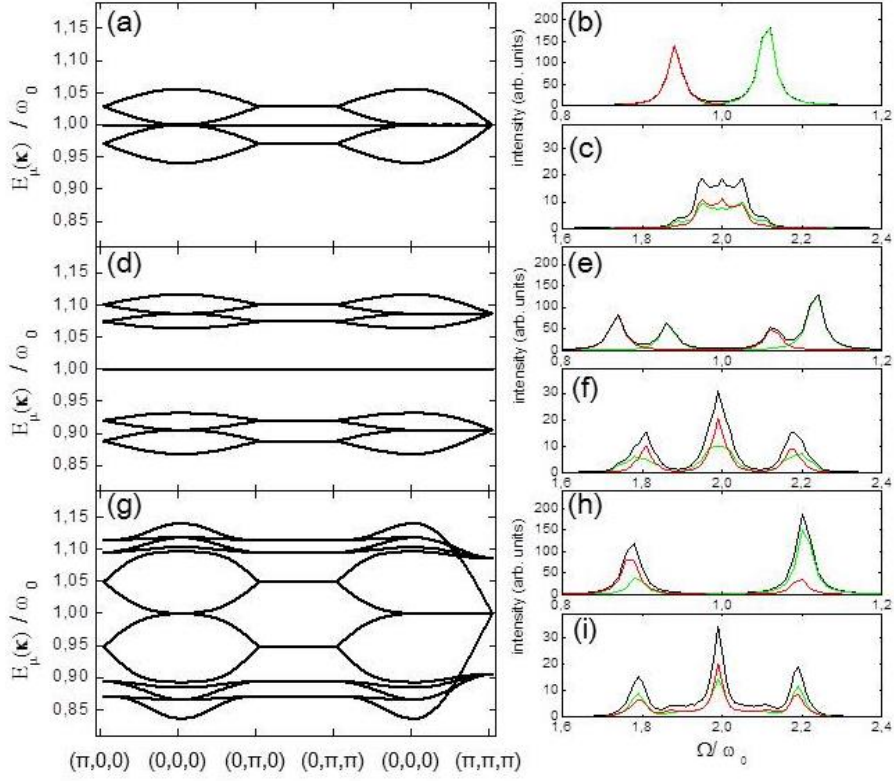

**FIG. 2** Dispersion curves [(a), (d), and (g)], the Raman scattering spectra near $\Omega/\omega_0 = 1.0$, i.e., the fundamental mode spectra [(b), (e), and (h)], and the Raman scattering spectra near $\Omega/\omega_0 = 2.0$, i.e, the first overtone mode spectra [(c), (f), and (i)] for different energy parameters. The energy parameters are $J/\omega_0 = 0.5$, $g/\omega_0 = 0$, and $\omega_1/\omega_0 = 0$ for (a), (b), and (c), respectively. $J/\omega_0 = 0.5$, $g/\omega_0 = 0.2$, and $\omega_1/\omega_0 = 0$ for (d), (e), and (f), respectively. $J/\omega_0 = 0.5$, $g/\omega_0 = 0.2$, and $\omega_1/\omega_0 = -0.1$ for (g), (h), and (i), respectively.

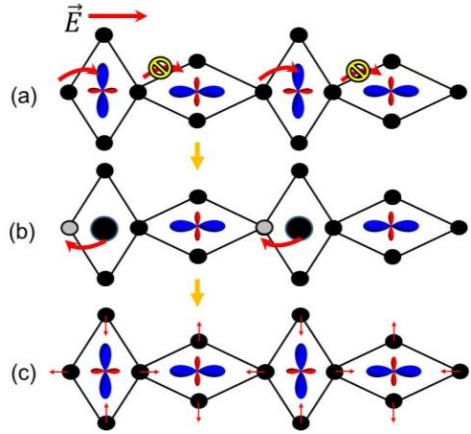

**FIG. 3** A schematic diagram for the photon absorption and the Raman mode generation for UV phonons of which energy is larger than approximately 3 eV. (a) Initial state: photon absorption causes the electron transition from oxygen ions to Mn ion. Because of the orbital shape, the electron at the specific oxygen can excite at this energy. (b) Intermediate state: the excited electrons move back to the original sites. Scattered photons are generated. (c) Final state: generation of a Raman mode in which energy corresponds to the energy difference between the absorbed photon and the scattered photon.